\documentclass[11pt]{article}

\usepackage{amsfonts,amsmath,amssymb,bm,graphicx,hyperref,a4wide,cite,mathabx,subfigure}

\begin{document}

\title{\textbf{Spin and flavor oscillations of neutrinos in gravitational fields}}

\author{Maxim Dvornikov\thanks{maxim.dvornikov@gmail.com} 
\\
\small{\ Pushkov Institute of Terrestrial Magnetism, Ionosphere} \\
\small{and Radiowave Propagation (IZMIRAN),} \\
\small{108840 Troitsk, Moscow, Russia}}

\date{}

\maketitle

\begin{abstract}
We summarize our recent achievements in the description of neutrino oscillations in various gravitational fields. After the short review of the previous studies of neutrinos in gravitational fields, we consider the neutrinos propagation and oscillations in two gravitational backgrounds. First, we discuss neutrino spin oscillations in their gravitational scattering off a supermassive black hole surrounded by a thick magnetized accretion disk. Second, we study neutrino flavor oscillations in stochastic gravitational waves. We also consider applications of the obtained results for oscillations of astrophysical neutrinos.
\end{abstract}


\section{Introduction}\label{sec:INTRO}

External fields can significantly modify neutrino oscillations making this process experimentally measurable. For instance, the neutrino electroweak interaction with background matter can amplify the transition probability of neutrino flavor oscillations~\cite{BleSmi13}. It is the most plausible explanation of the solar neutrinos deficit. The interaction with external electromagnetic field is crucial for neutrino spin and spin-flavor oscillations (see, e.g., Ref.~\cite{VolVysOku86}). The gravitational interaction, in spite of its weakness, can also contribute to neutrino oscillations.

The main problem in the description of the spinning particle evolution in a curved spacetime is the deviation of the particle trajectory from geodesics~\cite{Wal72}. However, this deviation was shown in Ref.~\cite{RieHol93} to be negligible in case of point-like elementary particles. Based on this result, the quasiclassical evolution equations for a spinning fermion in a curved spacetime were derived in Ref.~\cite{PomKhr98}. The quantum description of a spinning particle motion in a gravitational field was developed in Ref.~\cite{ObuSilTer17}.

The approach in Ref.~\cite{PomKhr98} was used in Ref.~\cite{Dvo06} to describe neutrino spin oscillations in a gravitational field. The validity of results in Ref.~\cite{Dvo06} was checked in Ref.~\cite{SorZil07}. The contributions of neutrino interactions with an electromagnetic field and a background matter were taken into account in this approach in Ref.~\cite{Dvo13}. The quasiclassical equation for a neutrino spin obtained in Ref.~\cite{Dvo13} were rederived basing on the Dirac equation in a curved spacetime in Ref.~\cite{Dvo19}.

The formalism for the description of neutrino spin oscillations developed in Refs.~\cite{Dvo06,Dvo13} was used in Ref.~\cite{AlaNod15} to study spin oscillations in gravitational fields in non-commutative spaces. This method was applied for the description of spin oscillations in some alternative gravity theories in Ref.~\cite{Cha15}. The suppression of neutrino spin oscillations because of the additional interaction with a scalar field was established in Ref.~\cite{MasLam21}. Neutrino spin oscillations in the vicinity of black holes described by the extensions of the General Relativity were considered in Ref.~\cite{Pan22}. Spin oscillations of neutrinos in curved spacetimes with higher dimensions were discussed in Ref.~\cite{Ala22}.

The influence of the gravitational interaction on neutrino flavor oscillations was studied in Refs.~\cite{For97,CarFul97}. Analogous approach was used in Ref.~\cite{PirRoyWud96} to describe neutrino spin-flavor oscillations in electromagnetic and gravitational fields. We also mention Ref.~\cite{Vic14} where neutrino flavor oscillations in various gravitational backgrounds were studied using the WKB approximation. Different approaches for the description of neutrino flavor oscillations in static metrics were compared in Ref.~\cite{GodPas11}.

Oscillations of neutrinos propagating in the vicinity of massive astrophysical objects should acquire gravity contributions. The gravitational lensing of supernova (SN) neutrinos of by a supermassive black hole (SMBH) was studied in Refs.~\cite{MenMocQui06,Los21}. Spin oscillations of such neutrinos were considered in Refs.~\cite{Dvo20a,Dvo20b,Dvo21,Dvo22} under the assumption that they propagate in the equatorial plane of a black hole (BH). The general situation, when the incoming neutrinos beam is both above and below the equatorial plane, was considered in Refs.~\cite{Dvo23c,Dvo23a,Dvo23d,Dvo23b}. These studies are inspired by the recent observations of the BH shadows in Refs.~\cite{Aki19,Aki22}.

Time dependent gravitational backgrounds also contribute to the fermion polarization. One of the most important examples for such gravitational fields is a gravitational wave (GW). We recall that GWs emitted by merging astrophysical objects are considered to be reliably detected (see, e.g., Ref.~\cite{Abb23}).

First, we mention Refs.~\cite{Qua16,ObuSilTer17} where the precession of the electron spin under the influence of GW was studied. The exact solution of the quasiclassical precession equation for the particle spin in GW was found in Ref.~\cite{PomSenKhr00}. The Bargmann-Michel-Telegdi equation in a curved spacetime corresponding to GW for a fermion possessing the anomalous magnetic moment was solved in Ref.~\cite{BalKurZim03}. Neutrino spin oscillations in matter under the influence of GW and a transverse magnetic field were studied in Ref.~\cite{Dvo19}. 

GWs can contribute to neutrino flavor oscillations as well. However, if GW propagates along the neutrino velocity, the impact of such GW on neutrino flavor oscillations was found in Ref.~\cite{Dvo19b} to be vanishing. Thus, unlike the case of an electromagnetic wave (see, e.g., Ref.~\cite{Dvo18}), the effects of multimessenger astronomy are non-observable for GWs. Nevertheless one can study the situation when a neutrino interacts with stochastic GWs~\cite{Ren22} emitted by randomly distributed sources. Neutrino flavor oscillations in GWs were also studied in Refs.~\cite{Dvo20c,KouMet20}.

The formalism, developed in Ref.~\cite{Dvo19b}, was used in Ref.~\cite{Dvo21b} to study the decoherence effects in flavor oscillations of supernova neutrinos. Stochastic GWs were supposed in Ref.~\cite{Dvo21b} to be produced by merging SMBHs. The importance of these studies is owing to the recent observations of stochastic GWs reported, e.g., in Refs.~\cite{Ant23,Xu23}. Note that the effects of decoherence in this case were shown in Ref.~\cite{Lam23} to be measurable in the JUNO detector. We also mention that the impact of relic stochastic GWs, produced before the electroweak phase transition, on neutrino flavor oscillations was obtained in Ref.~\cite{Dvo23e} to be negligible.

In the present work, we summarize our resent achievements on neutrino oscillations in various gravitational fields. This paper is organized in the following way. First, in Sec.~\ref{sec:SPIN}, we study neutrino spin oscillations in the gravitational scattering off a SMBH surrounded by a thick magnetized accretion disk. Then, in Sec.~\ref{sec:FLAV}, we consider neutrino flavor oscillations in stochastic GWs. Finally, we conclude in Sec.~\ref{sec:CONCL}.

\section{Spin oscillations in neutrino gravitational scattering\label{sec:SPIN}}

In this section, we review the formalism for the description of the neutrino spin evolution in a curved spacetime and apply it for the consideration of spin effects in the neutrino scattering off BH.

We suppose that that a neutrino is a Dirac fermion having nonzero magnetic moment $\mu$. If this neutrino interacts with the external electromagnetic field $F_{\mu\nu}$ and a background matter in a curved spacetime, its spin $S^\mu$ evolves as~\cite{Dvo13}
\begin{equation}\label{eq:Sevol}
  \frac{\mathrm{D}S^{\mu}}{\mathrm{d}\tau} = 2\mu
  \left(
    F^{\mu\nu}S_{\nu}-U^{\mu}U_{\nu}F^{\nu\lambda}S_{\lambda}
  \right)
  + \sqrt{2}G_{\mathrm{F}}E^{\mu\nu\lambda\rho}G_{\nu}U_{\lambda}S_{\rho},
\end{equation}
where $U^{\mu} = \tfrac{\mathrm{d}x^\mu}{\mathrm{d}\tau}$ is the neutrino four velocity in world coordinates $x^\mu$, $\tau$ is the proper time, $\mathrm{D}S^{\mu} = \mathrm{d}S^{\mu} + \Gamma^\mu_{\nu\lambda} S^\nu \mathrm{d}x^\lambda$ is the covariant differential, $G_{\mathrm{F}}$ is the Fermi constant, $E^{\mu\nu\lambda\rho} = \tfrac{1}{\sqrt{-g}}\varepsilon^{\mu\nu\lambda\rho}$ is the invariant antisymmetric tensor in curved spacetime, and $G^{\mu}$ is the four vector which incorporates the matter contribution. The explicit form of this vector is given in Refs.~\cite{Dvo13,DvoStu02}. We should add the geodesics equation,
\begin{equation}
  \frac{\mathrm{D}U^{\mu}}{\mathrm{d}\tau} = 0,
\end{equation}
to Eq.~\eqref{eq:Sevol} to account for the action of the gravitational field on the particle trajectory.

The invariant polarization is described in the particle rest frame which is defined in the locally Minkowskian frame $x^a = e^a_{\,\mu}x^\mu$, where $e^a_{\,\mu}$ are the vierbein vectors. These vector satisfy the relation, $\eta_{ab} = e_a^{\,\mu}e_b^{\,\nu}g_{\mu\nu}$, where $\eta_{ab} = \text{diag}(+1,-1,-1,-1)$ is the flat spacetime metric tensor.

The invariant three vector of the neutrino polarization $\bm{\bm{\zeta}}$ obeys the equation,
\begin{equation}\label{eq:nuspinrot}
  \frac{\mathrm{d}\bm{\bm{\zeta}}}{\mathrm{d}t}=2(\bm{\bm{\zeta}}\times\bm{\bm{\Omega}}),
  \quad
  \bm{\bm{\Omega}}=\bm{\bm{\Omega}}_{g}+\bm{\bm{\Omega}}_{\mathrm{em}}+\bm{\bm{\Omega}}_{\mathrm{matt}},
\end{equation}
where $\bm{\bm{\Omega}}_{g,\mathrm{em,matt}}$ are the contributions of the neutrino interactions with gravitational and electromagnetic fields, as well as the electroweak interaction with matter. The explicit forms of these vectors are given in Refs.~\cite{Dvo23c,Dvo23a,Dvo23b}. 

We apply this formalism for the studies of the neutrino motion and spin oscillations in the gravitational fields of a rotating BH surrounded by a thick magnetized accretion disk. If we use the Boyer-Lindquist coordinates $x^{\mu}=(t,r,\theta,\phi)$, the spacetime metric has the form,
\begin{equation}\label{eq:Kerrmetr}
  \mathrm{d}\tau^{2} = g_{\mu\nu}\mathrm{d}x^{\mu}\mathrm{d}x^{\nu}=
  \left(
    1-\frac{rr_{g}}{\Sigma}
  \right)
  \mathrm{d}t^{2}
  + 2\frac{rr_{g}a\sin^{2}\theta}{\Sigma}\mathrm{d}t\mathrm{d}\phi-\frac{\Sigma}{\Delta}\mathrm{d}r^{2}
  - \Sigma\mathrm{d}\theta^{2} -
  \frac{\Xi}{\Sigma}\sin^{2}\theta\mathrm{d}\phi^{2},
\end{equation}
where
\begin{align}\label{eq:dsxi}
  \Delta= & r^{2}-rr_{g}+a^{2},
  \quad
  \Sigma=r^{2}+a^{2}\cos^{2}\theta,
  \notag
  \\
  \Xi= & 
  \left(
    r^{2}+a^{2}
  \right)
  \Sigma+rr_{g}a^{2}\sin^{2}\theta.
\end{align}
In Eqs.~\eqref{eq:Kerrmetr} and~\eqref{eq:dsxi}, the mass of BH is $M$, its angular momentum, which along the $z$-axis, is $J = aM$, and  $r_g = 2M$ is the Schwarzschild radius.

The equations of motion of a test particle in metric in Eqs.~\eqref{eq:Kerrmetr} and~\eqref{eq:dsxi} were shown in Ref.~\cite{GraLupStr18} to be integrated in quadratures. One has three conserved quantities in case of ultrarelativistic neutrinos: the particle energy $E$, the projection of its angular momentum on the BH rotation axis $L$, and the Carter constant $Q$. In the following, we use the dimensionless variables, $r=xr_{g}$, $L=yr_{g}E$, $Q=wr_{g}^{2}E^{2}$, and $a=zr_{g}$. The detailed description of the neutrino motion in the Kerr metric is provided in Refs.~\cite{Dvo23c,Dvo23a,Dvo23b}.

Instead of Eq.~\eqref{eq:nuspinrot}, we deal with the effective Schr\"odinger equation when we describe the neutrino spin evolution in the Kerr metric,
\begin{align}\label{eq:Schreq}
  \mathrm{i}\frac{\mathrm{d}\psi}{\mathrm{d}x}= & \hat{H}_{x}\psi,
  \quad
  \hat{H}_{x}=-\mathcal{U}_{2}(\bm{\bm{\sigma}}\cdot\bm{\bm{\Omega}}_{x})\mathcal{U}_{2}^{\dagger},
  \notag
  \\
  \bm{\bm{\Omega}}_{x} = & r_{g}\bm{\bm{\Omega}}\frac{\mathrm{d}t}{\mathrm{d}r},
  \quad
  \mathcal{U}_{2}=\exp(\mathrm{i}\pi\sigma_{2}/4).
\end{align}
Here, $\bm{\bm{\sigma}}=(\sigma_{1},\sigma_{2},\sigma_{3})$ are the Pauli matrices. The Hamiltonian $\hat{H}_{x}$ is the function of $x$ only since the dependence $\theta(x)$ results from the particle trajectory. The initial condition for Eq.~(\ref{eq:Schreq}) has the form, $\psi_{-\infty}^{\mathrm{T}}=(1,0)$, which corresponds to left polarized incoming neutrinos. The polarization of an outgoing neutrino is represented in the form, $\psi_{+\infty}^{\mathrm{T}}=(\psi_{+\infty}^{(\mathrm{R})},\psi_{+\infty}^{(\mathrm{L})})$.
The survival probability is $P_{\mathrm{LL}}=|\psi_{+\infty}^{(\mathrm{L})}|^{2}$. 

We suppose that SMBH with $M=10^8 M_\odot$ is surrounded by a thick magnetized accretion disk, which is described in frames of the Polish doughnut model~\cite{AbrJarSik78,Kom06}. This disk is taken to consist of hydrogen plasma, with the maximal electron number density being equal to $10^{18}\,\text{cm}^{-3}$, which is not observationally excluded~\cite{Jia19}. The considered model of the accretion disk implies the presence of a toroidal magnetic field. We suppose that the maximal strength of this field is $\sim 320\,\text{G}$. Such a strength is $\sim 1\%$ of the Eddington limit~\cite{Bes10}, that is quite moderate value. We also assume that the neutrino magnetic moment is $\mu = 10^{-13}\mu_\mathrm{B}$, that is below the astrophysical constraint in Ref.~\cite{Via13}.

Since we can observe only left-polarized neutrinos, the measurable flux is $F_\nu = P_{\mathrm{LL}} F_0$, where $F_0$ is the flux of spinless particles, which move along geodesics only. Thus we should find $F_\nu/F_0$ for outgoing neutrinos.

First, we calculated $P_{\mathrm{LL}}$ for purely gravitational interaction, when the accretion disk is absent. One get that $P_{\mathrm{LL}}= 1$ with high level of accuracy. It means that the helicity of ultrarelativistic neutrinos is conserved in their scattering off a rotating black hole. This fact generalizes the analogous finding in Ref.~\cite{Lam05}, which was established in the weak gravitational field approximation. We also found that the toroidal magnetic field in frames of the model in Ref.~\cite{Kom06} does not contribute to neutrino spin oscillations for any reasonable strengths of this field. The explanation of this fact is given in Ref.~\cite{Dvo23b}.

Thus, to get any spin oscillations in this system, we have to consider a poloidal magnetic field. It was shown in Ref.~\cite{BraNot06}, that only the configuration of magnetic field which contains both toroidal and poloidal components is stable. We discuss two models of the poloidal magnetic field: (i) that in Ref.~\cite{Wal74}; and (ii) the poloidal field in Ref.~\cite{FraMei09}. The maximal strength of the poloidal magnetic field is chosen to be $\sim 320\,\text{G}$, i.e. it is equal to that of the toroidal one.

\begin{figure*}[htbp]
  \centering
  \subfigure[]
  {\label{fig:f1a}
  \includegraphics[scale=.35]{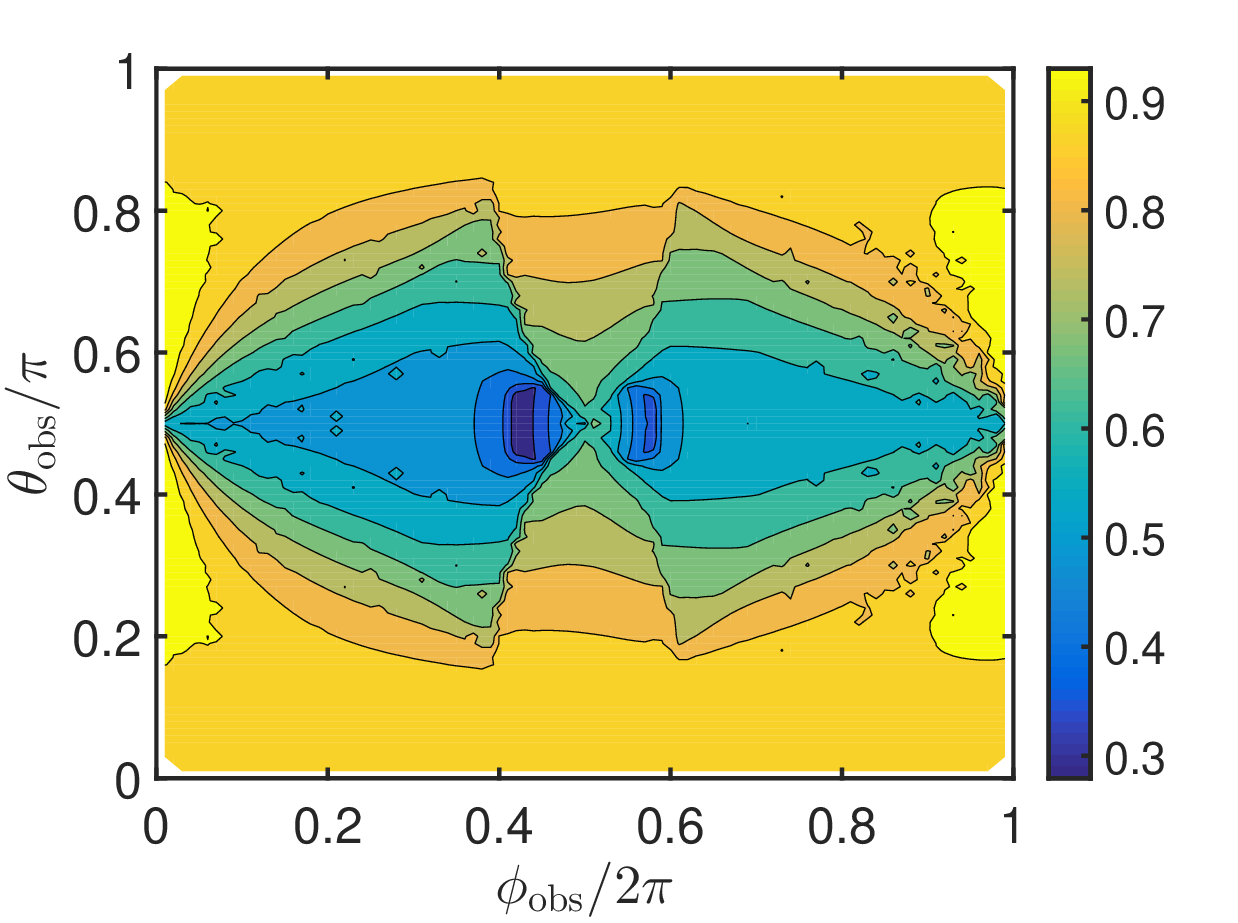}}
  \hskip-.5cm
  \subfigure[]
  {\label{fig:f1b}
  \includegraphics[scale=.35]{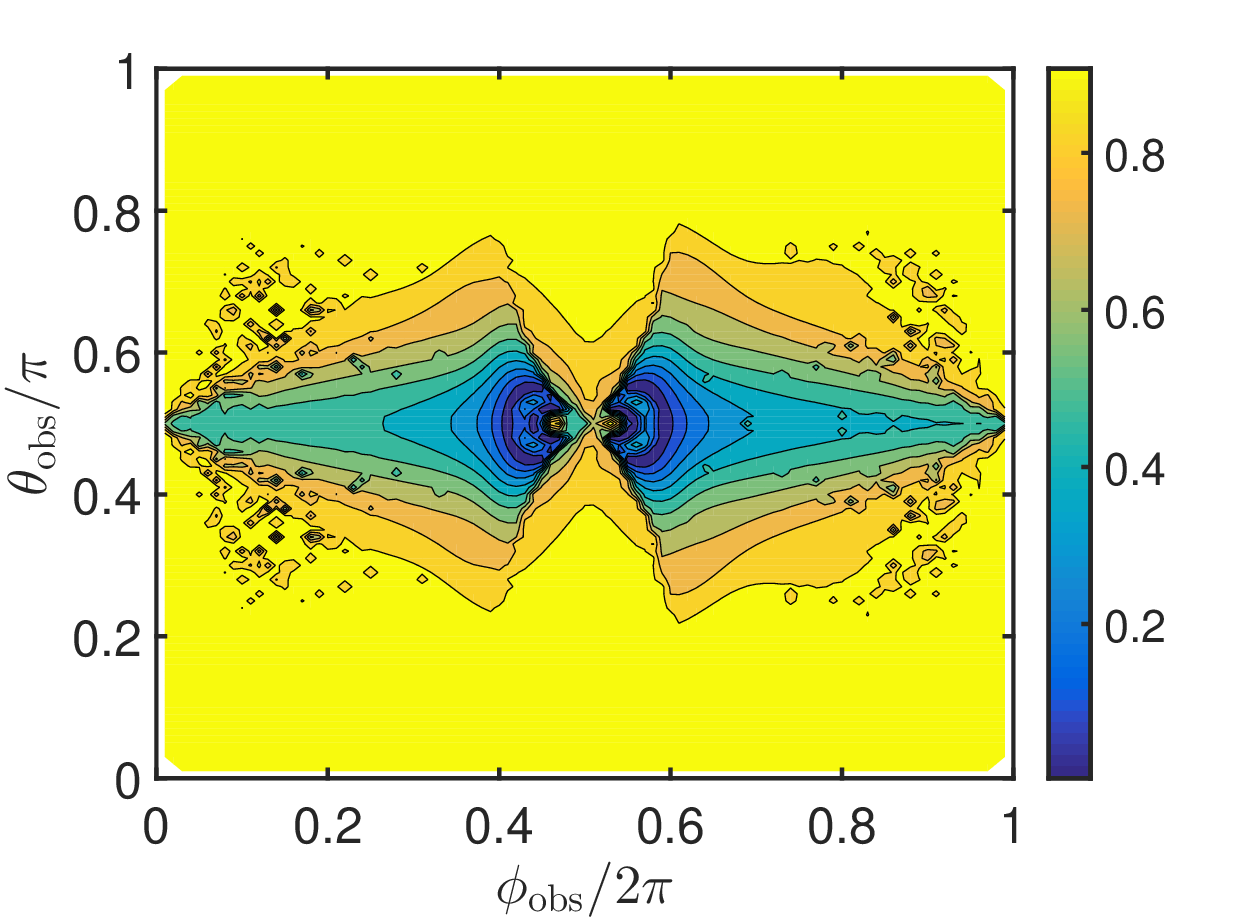}}
  \\
  \subfigure[]
  {\label{fig:f1c}
  \includegraphics[scale=.35]{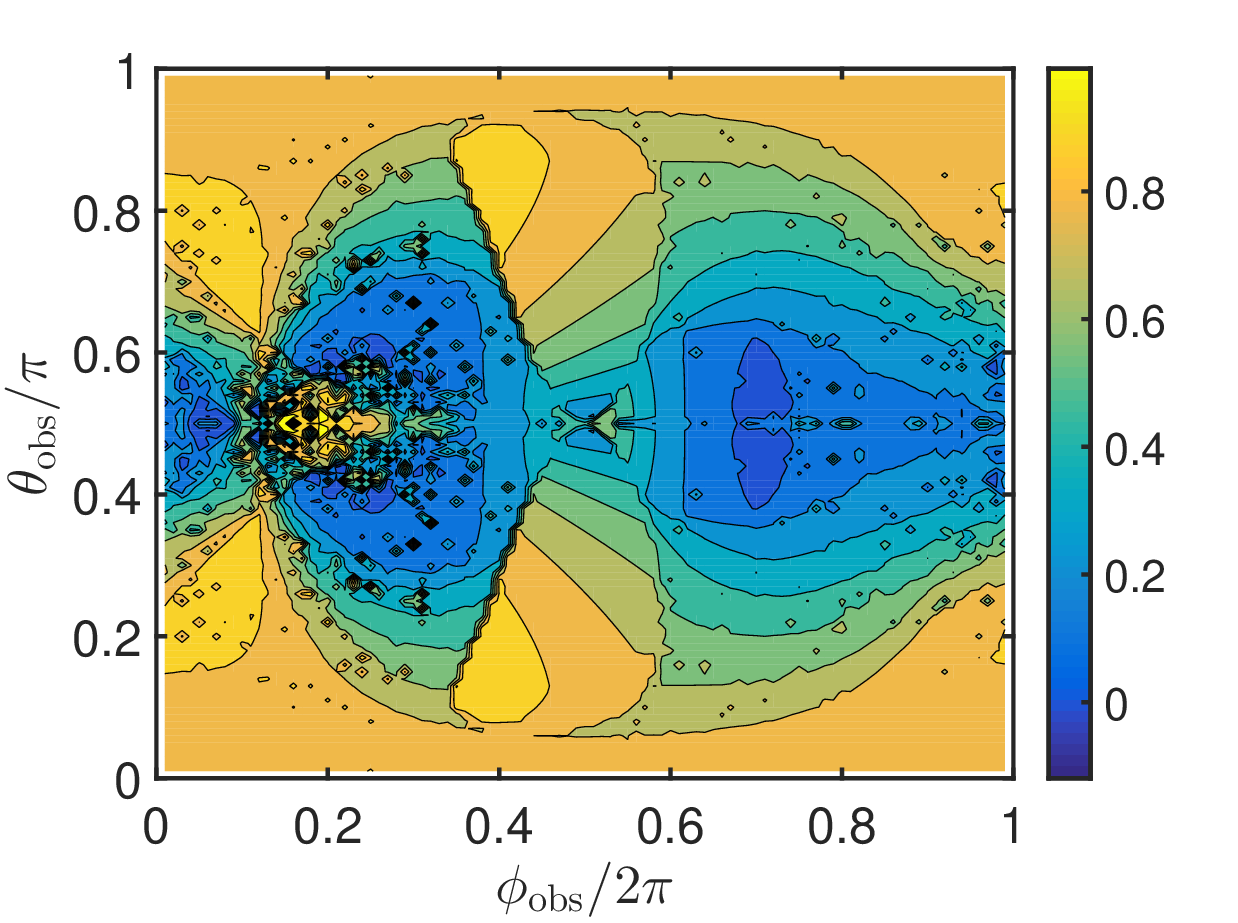}}
  \hskip-.5cm
  \subfigure[]
  {\label{fig:f1d}
  \includegraphics[scale=.35]{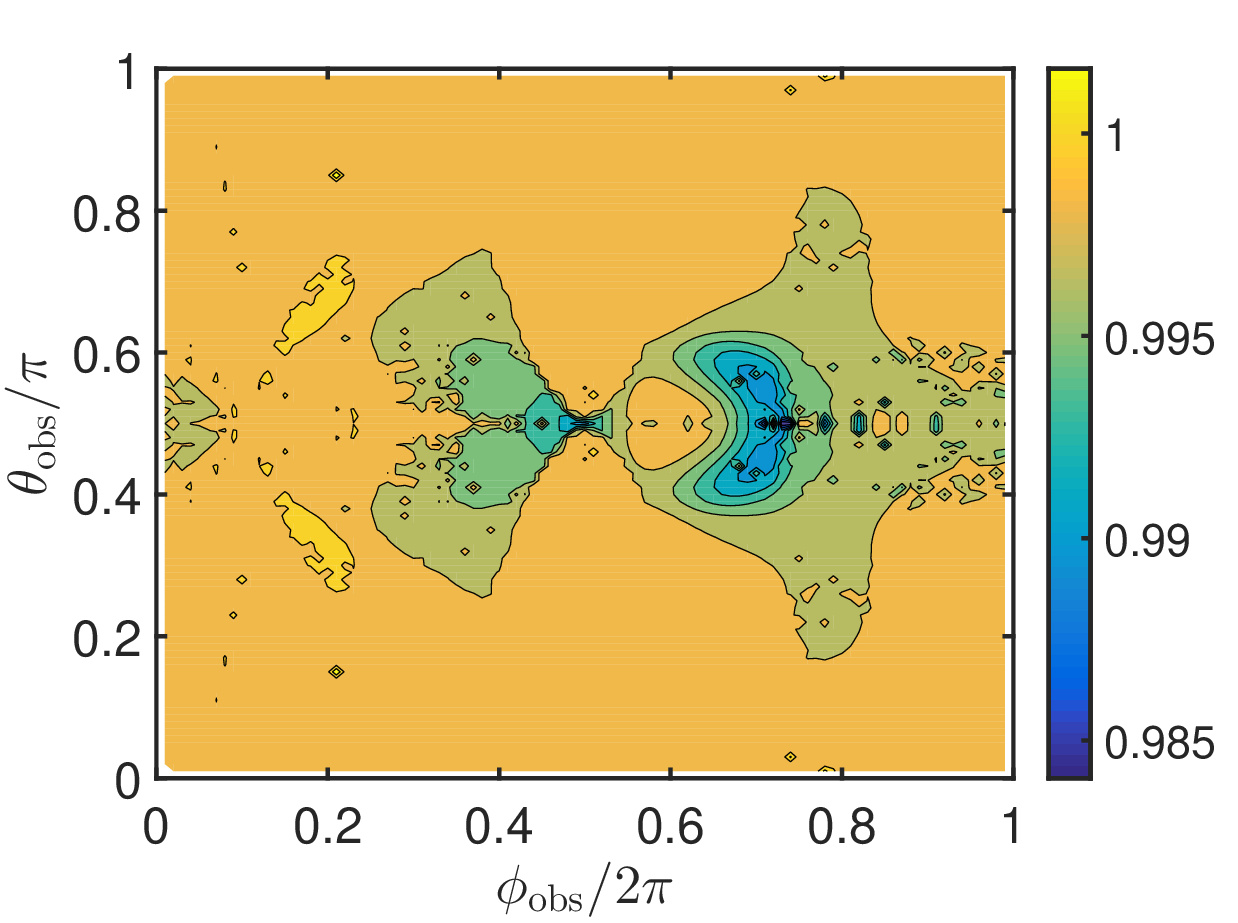}}
  \protect 
\caption{The flux of spinning neutrinos $F_\nu$ normalized by the flux of scalar particles $F_0$ for various models of the poloidal magnetic field and spins of BH. Panels~(a) and~(b) are built for $a = 2\times 10^{-2} M$; panels~(c) and~(d) -- for $a = 0.9 M$. Panels~(a) and~(c) correspond to the poloidal field in Ref.~\cite{Wal74}; panels~(c) and~(d) -- in Ref.~\cite{FraMei09}.\label{fig:flux}}
\end{figure*}

In Fig.~\ref{fig:flux}, we show the observed flux of neutrinos for different models of poloidal fields and various spins of SMBH. In Fig.~\ref{fig:flux}, the image of BH is at $\phi_\mathrm{obs} = \pi$ and $\theta_\mathrm{obs} = \pi/2$. Looking at the neutrino fluxes in Figs.~\ref{fig:f1a} and~\ref{fig:f1b}, which correspond to slowly rotating BHs, one can identify the distribution of poloidal magnetic fields in the accretion disk. If the spin of BH increases, the manifestation of spin effects in the neutrino scattering is not so straightforward; cf. Figs.~\ref{fig:f1c} and~\ref{fig:f1d}. Moreover, as one can see in Fig.~\ref{fig:f1d}, neutrino spin oscillations are almost vanishing in the model in Ref.~\cite{FraMei09}. It happens since the poloidal field is nonzero within a thin torus near the BH surface.

The results in Fig.~\ref{fig:flux} demonstrate the possibility of the neutrino tomography of the magnetic fields distribution in the vicinity of BH. The analysis in Ref.~\cite{Dvo23c} shows that spin effects in gravitational scattering of SN neutrinos are potentially measurable in future neutrino detectors like the Hyper-Kamiokande~\cite{Abe21}.

\section{Spin oscillations in stochastic GWs\label{sec:FLAV}}

In this section, we describe neutrino flavor oscillations in a plane GW with arbitrary polarizations and apply the results for the description of decoherence effects in oscillations of SN neutrinos in stochastic GWs.

We suppose that we deal with three active neutrinos $\nu_\lambda$, $\lambda = e,\mu,\tau$, which are the superposition of the mass eigenstates $\psi_a$, $a=1,2,3$: $\nu_\lambda = (U)_{\lambda a} \psi_a$, where $(U)_{\lambda a}$ are the components of the mixing matrix. The wave function of a neutrino mass eigenstate $\psi_a(\mathbf{x},t)$ evolves in a spacetime as $\psi_a(\mathbf{x},t) = \exp[ -\mathrm{i} S_a(\mathbf{x},t) ] \psi_a^{(0)}$, where $S_a(\mathbf{x},t)$ is the action corresponding to this eigenstate. In a spacetime with a given metric $g_{\mu\nu}$, $S_a(\mathbf{x},t)$ obeys the Hamilton-Jacobi equation,
\begin{equation}\label{eq:HJeq}
  g_{\mu\nu}\frac{\partial S_{a}}{\partial x_{\mu}}\frac{\partial S_{a}}{\partial x_{\nu}}=m_{a}^{2}.
\end{equation}
where $m_a$ is the mass of the eigenstate $\psi_a$.

We take that the metric of the spacetime corresponds to a plane GW, which propagates along the $z$-axis, with two independent polarizations,
\begin{equation}\label{eq:metric}
  \mathrm{d}s^{2}=
  g_{\mu\nu}\mathrm{d}x^{\mu}\mathrm{d}x^{\nu}=
  \mathrm{d}t^{2}
  - \left(
    1-h_+\cos\phi
  \right)
  \mathrm{d}x^{2} -
  \left(
    1+h_+\cos\phi
  \right)
  \mathrm{d}y^{2}
  + 2h_\times\sin\phi \, \mathrm{d}x\mathrm{d}y-\mathrm{d}z^{2},
\end{equation}
where $h_{+,\times}$ are the amplitudes of `plus' and `times' polarizations, $\phi = \omega (t - z)$, and $\omega$ is the frequency of GW.

The exact solution of Eq.~\eqref{eq:HJeq} with 
the metric in Eq.~\eqref{eq:metric} was found in Ref.~\cite{Pop06}. Instead of analyzing Eq.~\eqref{eq:HJeq}, we can find the contribution of GW to the effective Hamiltonian for neutrino flavor oscillations~\cite{Dvo21b},
$H = U H_m U^\dag$, where $H_m = H_m^{(\mathrm{vac})} + H_m^{(g)}$ is the Hamiltonian for mass eigenstates, $H_m^{(\mathrm{vac})} = \tfrac{1}{2E}\text{diag}\left(0,\Delta m_{21}^{2},\Delta m_{31}^{2}\right)$ is the part corresponding to vacuum oscillations, $\Delta m_{ab}^{2}=m_{a}^{2}-m_{b}^{2}$ is the difference of the squares of masses, and $E$ is the mean neutrino energy in a beam.

The GW contribution to the Hamiltonian has the form~\cite{Dvo21b},
\begin{align}\label{eq:Hgmass}
  H_{m}^{(g)} = & H_{m}^{(\mathrm{vac})}\left(A_{c}h_{+}+A_{s}h_{\times}\right),
  \notag
  \\
  A_{c,s} = & \frac{1}{2}\sin^{2}\vartheta \times
  \begin{cases}
    \cos2\varphi\cos[\omega t(1-\cos\vartheta)], \\
    \sin2\varphi \sin[\omega t(1-\cos\vartheta)], \\
  \end{cases}
\end{align}
where $\vartheta$ and $\varphi$ are the spherical angles fixing the neutrino velocity with respect to the GW wavevector. It should be noted that, if $\vartheta = 0$, i.e. a neutrino propagates along GW, $H_{m}^{(g)}$ in Eq.~\eqref{eq:Hgmass} vanishes. It means that the effects of the multimessenger astronomy, when GWs and neutrinos are emitted by the same source, are unachievable here.

Nevertheless, we can consider the situation, when a neutrino interacts with stochastic GWs emitted by randomly distributed sources, while a particle propagates from a source to a detector. In this case, it is more convenient to deal with the quantum Liouville equation for the density matrix $\rho$. If we define $\rho' = U^\dag \exp(\mathrm{i} H_0 t) \rho  \exp(-\mathrm{i} H_0 t) U$, where $H_0 = U H_m^{(\mathrm{vac})} U^\dag$, the evolution equation for $\rho'$ reads~\cite{Dvo21b},
\begin{align}\label{eq:rho'eq}
  \dot{\rho}' = & -g[H_{m}^{(\mathrm{vac})},[H_{m}^{(\mathrm{vac})},\rho']],
  \notag
  \displaybreak[2]
  \\
  g(t) = & \frac{3}{128}\int_{0}^{t}\mathrm{d}t_{1}
  \big(
    \left\langle
      h_{+}(t)h_{+}(t_{1})
    \right\rangle 
    + \left\langle
      h_{\times}(t)h_{\times}(t_{1})
    \right\rangle
  \big).
\end{align}
Equation~\eqref{eq:rho'eq} is integrated analytically and  the probabilities to observe the certain neutrino flavors in the neutrino beam, traveled the distance $x\approx t$, can be obtained.

However, the main contribution to the probabilities results from the vacuum oscillations term $P_{\lambda}^{(\mathrm{vac})}$ (see, e.g., Ref.~\cite{GiuKim07}).  If we subtract $P_{\lambda}^{(\mathrm{vac})}$ from the total probability, we get the contribution of GWs only in the form,
\begin{align}\label{eq:DeltaPgen}
  \Delta P_{\lambda}(x)= & 2\sum_{\sigma}P_{\sigma}(0)
  \sum_{a>b}
  \bigg\{
    \text{Re}
    \left[
      U_{\lambda a}U_{\lambda b}^{*}U_{\sigma a}^{*}U_{\sigma b}
    \right]
    \cos
    \left(
      2\pi\frac{x}{L_{ab}}
    \right)
    \notag
    \displaybreak[2]
    \\
    & +
    \text{Im}
    \left[
      U_{\lambda a}U_{\lambda b}^{*}U_{\sigma a}^{*}U_{\sigma b}
    \right]
    \sin
    \left(
      2\pi\frac{x}{L_{ab}}
    \right)
  \bigg\}
  \left\{
    1-\exp
    \left[
      -\frac{4\pi^{2}}{L_{ab}^{2}}\int_{0}^{x}g(t)\mathrm{d}t
    \right]
  \right\},
\end{align}
where $L_{ab}=\tfrac{4\pi E}{|\Delta m_{ab}^{2}|}$ are the neutrino oscillations lengths in vacuum and $P_\sigma(0)$ are the initial probabilities at $x=0$ which satisfy $\sum_\sigma P_\sigma(0) = 1$.

We suppose that neutrinos are emitted by a core-collapsing SN during the neutronization burst. In this case, the size of the neutrinosphere is $\lesssim 100\,\text{km}$. Thus, only the solar oscillations channel survives in Eq.~\eqref{eq:DeltaPgen} after averaging over neutrino emission positions. The ratio of the neutrino fluxes at a source is $\left(F_{\nu_{e}}:F_{\nu_{\mu}}:F_{\nu_{\tau}}\right)_{\mathrm{S}}=(1:0:0)$.

If we consider GWs of the astrophysical origin, the main contribution to the GWs background is taken to result from merging SMBHs. The spectrum of the energy density in this case was calculated in Ref.~\cite{Ros11}. Note that recent observations of stochastic GWs in Refs.~\cite{Ant23,Xu23} also indicate that they are likely to be produced by coalescing SMBHs. Using the results of Ref.~\cite{Dvo21b}, we calculate the fluxes of all flavor neutrinos  from SN, as well as the decoherence coefficient,
\begin{equation}\label{eq:DeltaP21}
  \Gamma= \frac{4\pi^{2}}{L_{21}^{2}}\int_{0}^{x}g(t)\mathrm{d}t,
\end{equation}
as functions of the distance $x$. We take into account only the solar oscillations channel in Eq.~\eqref{eq:DeltaP21}.

\begin{figure*}
  \centering
  \subfigure[]
  {\label{2a}
  \includegraphics[scale=.35]{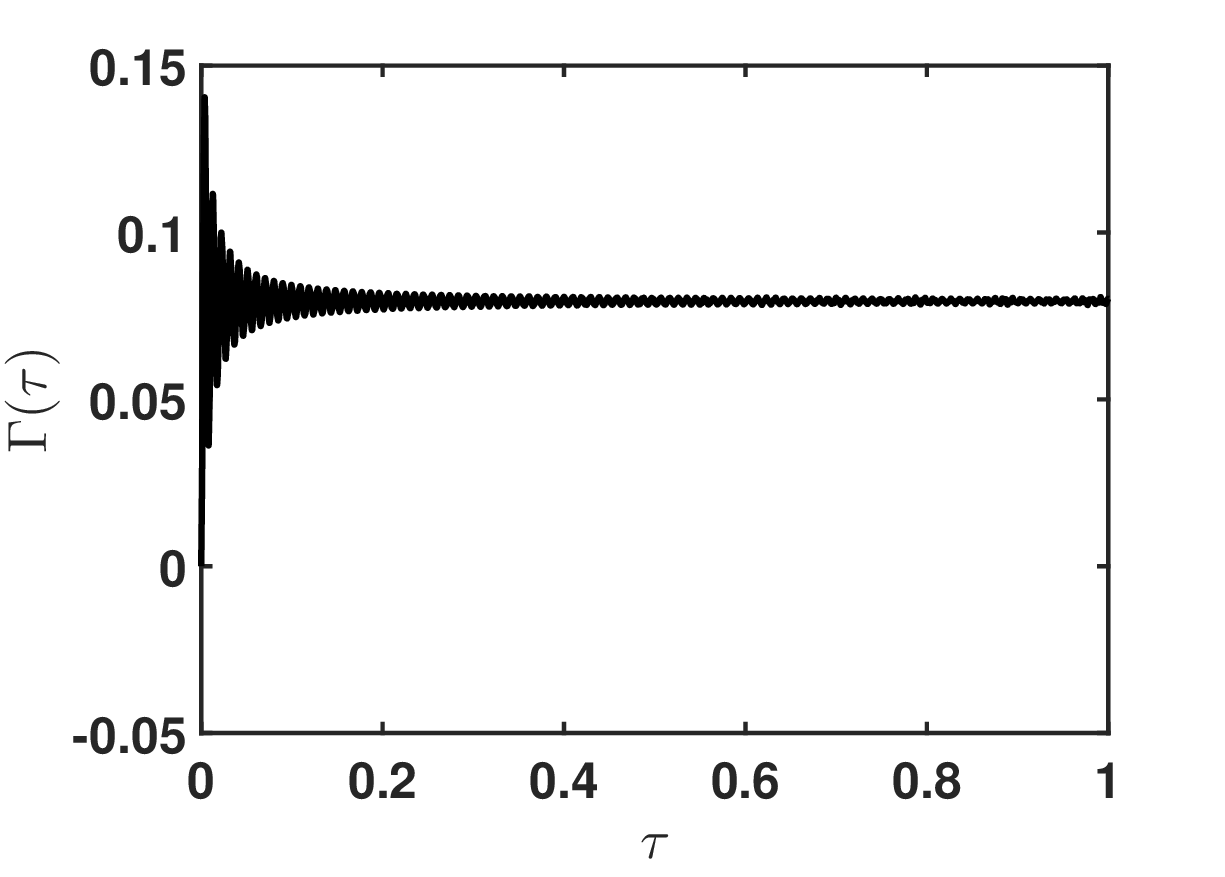}}
  \hskip-.6cm
  \subfigure[]
  {\label{2b}
  \includegraphics[scale=.35]{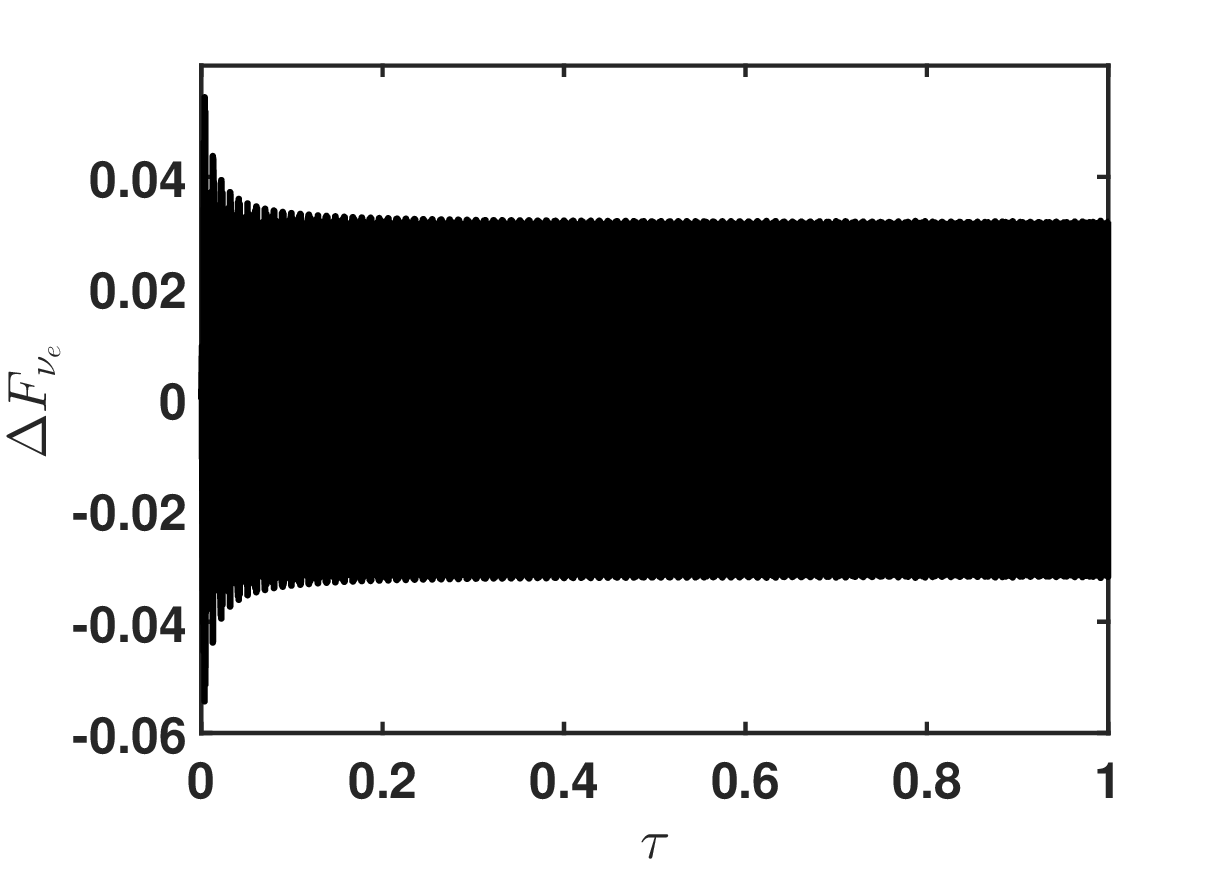}}
  \\
  \subfigure[]
  {\label{2c}
  \includegraphics[scale=.35]{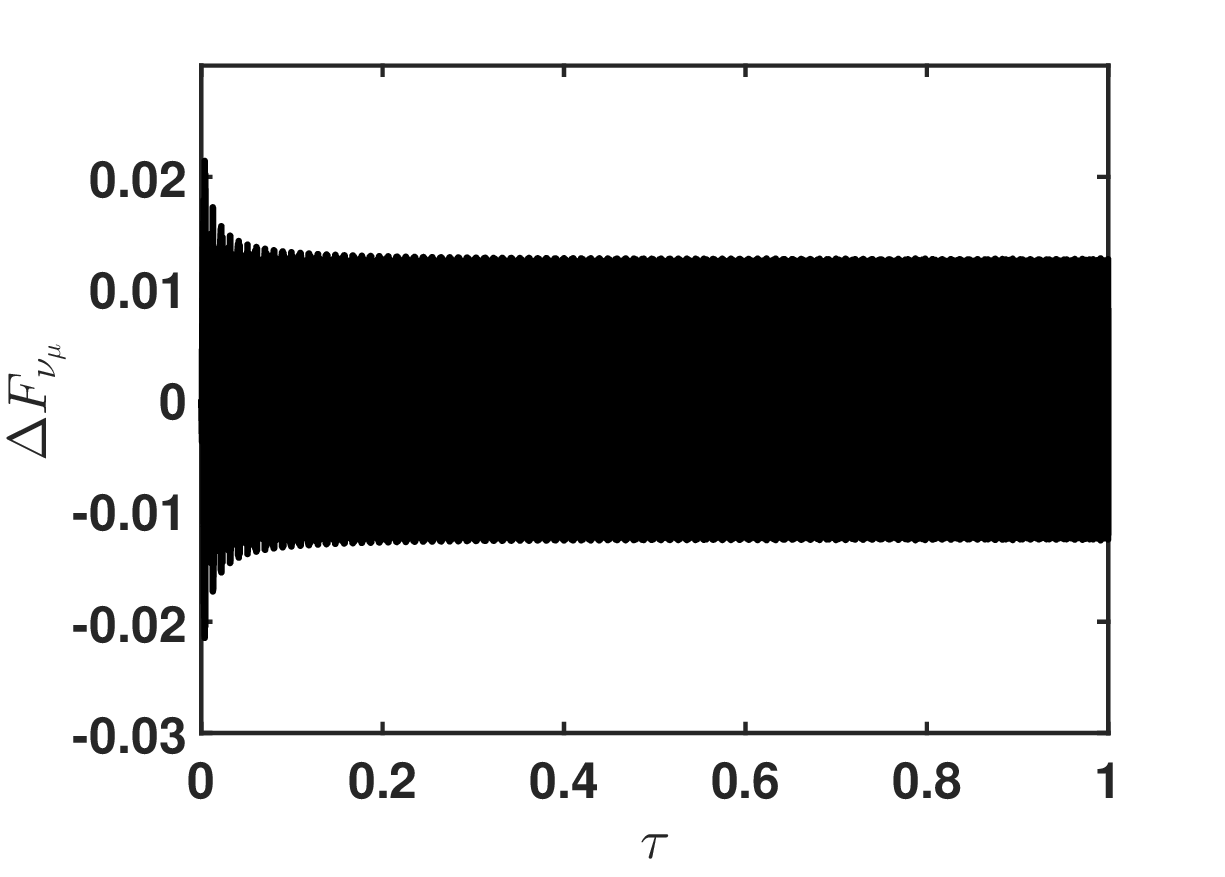}}
  \hskip-.6cm
  \subfigure[]
  {\label{2d}
  \includegraphics[scale=.35]{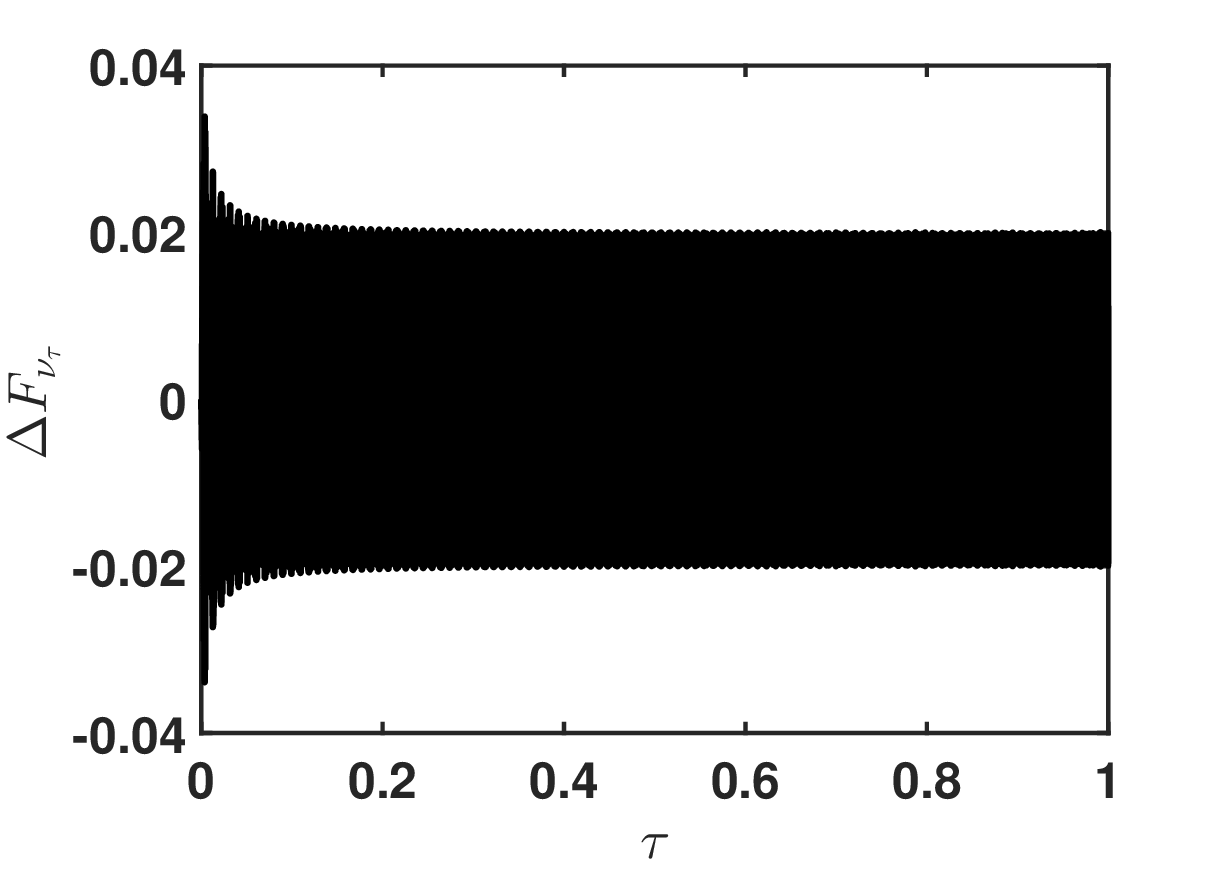}}
  \protect
  \caption{(a) The decoherence parameter $\Gamma$ in Eq.~\eqref{eq:DeltaP21} versus the propagation distance of a neutrino beam $\tau = x/ L$
  normalized by $L = 10\,\text{kpc}$; (b)-(d) the corrections to the neutrino fluxes $\Delta F_{\nu_{\lambda}}\propto\Delta P_{\lambda}$
  owing to the neutrino interaction with stochastic GWs. The figure is taken from Ref.~\cite{Dvo21b}.\label{fig:deltaF}}
\end{figure*}

In Fig.~\ref{fig:deltaF}, we show the corrections to the neutrino fluxes owing to the interaction with stochastic GWs. The parameters of neutrinos are~\cite{Sal20} $\Delta m_{21}^{2}=7.5\times10^{-5}\,\text{eV}^{2}$, $\theta_{12}=0.6$, $\theta_{23}=0.85$, $\theta_{13}=0.15$, and $\delta_{\mathrm{CP}}=3.77$. We assume the normal neutrino mass ordering. The energy of SN neutrinos is ~\cite{VitTamRaf20} $E=10\,\text{MeV}$. The characteristics of the spectrum of the energy density of GWs correspond to those calculated in Ref.~\cite{Ros11}.

One can see in Fig.~\ref{2a} that $\Gamma$ reaches its asymptotic value at $x\sim 1\,\text{kpc}$. Thus the decoherence effects are observable if a SN explosion takes place in our Galaxy. More detailed analysis in Ref.~\cite{Lam23}, based on results of observations in Refs.~\cite{Ant23,Xu23}, shows that the decoherence of SN neutrinos are measurable in JUNO detector~\cite{Abu22}. One can see in Figs.~\ref{2b}-\ref{2d} that the deviation of the fluxes from the vacuum oscillations values can reach $(1-3)\%$. In case when a SN explosion happens in our Galaxy and SN neutrinos are measured, e.g., with the Hyper-Kamiokande detector~\cite{Abe21}, the deviation of the fluxes owing the neutrino interaction with stochastic GWs can reach several thousand events.

\section{Conclusion\label{sec:CONCL}}

In the present work, we have summarized our recent results on the description of propagation and oscillations of neutrinos in gravitational fields of various configurations. We have started in Sec.~\ref{sec:INTRO} with a brief review on previous works where neutrino oscillations in gravitational fields were discussed. We have mentioned mainly the works on spin and flavor oscillations.

Then, in Sec.~\ref{sec:SPIN}, we have considered spin oscillations of neutrinos gravitationally scattered off SMBH surrounded by a magnetized accretion disk. The trajectories of neutrinos have been described exactly in our treatment. We have accounted for the neutrino spin interaction with gravitational and magnetic fields, as well as the electroweak interaction with plasma of the disk. We have found that the spin-flip does not occur in case ultrarelativistic  neutrinos participating in only the gravitational interaction, i.e. at the absence of a disk. The nonzero contribution to neutrino spin oscillations results from the poloidal magnetic field. We have obtained the fluxes of spinning neutrinos for different angular momenta of BH and various models of poloidal magnetic fields.

We have studied neutrino flavor oscillations in the time dependent gravitational background which corresponds to GW. We have started with consideration of a plane GW with two independent polarizations and derived the contribution to the effective Hamiltonian for flavor oscillations. Then, we have studied the situation of stochastic GWs and found the expression for the neutrino density matrix in the exact form. Based on this result, we have obtained the probabilities for all neutrino flavors. Finally, we have discussed applications for the decoherence of SN neutrinos in the GW background emitted by coalescing SMBHs.

%
%


\nocite{*}


\end{document}